\documentclass{jpp}

\usepackage{color}
\usepackage{graphicx}
\usepackage{natbib}

\ifCUPmtlplainloaded \else
  \checkfont{msam10}
  \iffontfound
    \IfFileExists{amssymb.sty}
      {\typeout{^^JFound AMS Symbol fonts on the system, using the
                'amssymb' package.^^J}%
       \usepackage{amssymb}%
         
         \let\geq=\geqslant
      }{}
  \fi
\fi

\ifCUPmtlplainloaded \else
  \IfFileExists{amsbsy.sty}
    {\typeout{^^JFound the 'amsbsy' package on the system, using it.^^J}%
     \usepackage{amsbsy}}
    {}
\fi

\newsavebox{\astrutbox}
\sbox{\astrutbox}{\rule[-5pt]{0pt}{20pt}}

\title[Particle trajectories in Weibel filaments]{Particle trajectories in Weibel filaments: influence of external field obliquity and chaos}

\author[A. Bret and M. E. Dieckmann]%
{A.  Bret$^{1,2}$, and M. E. Dieckmann$^3$%
  \thanks{Email address for correspondence: antoineclaude.bret@uclm.es}
}

\affiliation{$^1$ETSI Industriales, Universidad de Castilla-La Mancha, 13071 Ciudad Real, Spain\\[\affilskip]
$^2$Instituto de Investigaciones Energ\'{e}ticas y Aplicaciones Industriales, Campus Universitario de Ciudad Real, 13071 Ciudad Real, Spain\\[\affilskip]
$^3$Department of Science and Technology (ITN), Link\"{o}ping University, 60174 Norrk\"{o}ping, Sweden
}

\date{?; revised ?; accepted ?. - To be entered by editorial office}

\begin{document}

\maketitle

\begin{abstract}
When two collisionless plasma shells collide, they interpenetrate and the overlapping region may turn Weibel unstable for some values of the collision parameters. This instability grows magnetic filaments which, at saturation, have to block the incoming flow if a Weibel shock is to form. In a recent paper [J. Plasma Phys. (2016), vol. 82, 905820403], it was found implementing a toy model for the incoming particles trajectories in the filaments, that a strong enough external magnetic field $\mathbf{B}_0$ can prevent the filaments to block the flow if it is aligned with. Denoting $B_f$ the peak value of the field in the magnetic filaments, all test particles stream through them if $\alpha=B_0/B_f > 1/2$. Here, this result is extended to the case of an oblique external field $B_0$ making an angle $\theta$ with the flow. The result, numerically found, is simply $\alpha > \kappa(\theta)/\cos\theta$, where $\kappa(\theta)$ is of order unity. Noteworthily, test particles exhibit chaotic trajectories.
\end{abstract}

\maketitle

\section{Introduction}

Collisionless plasmas can sustain shock waves with a front much smaller than the particles mean-free-path \citep{Sagdeev_Kennel_1991}. These shocks, which are mediated by collective plasma effects rather than binary collisions, have been dubbed ``collisionless shocks''. It is well known that the encounter of two collisional fluids generates two counter-propagating shock waves \citep{Zeldovich}. Likewise, the encounter of two collisionless plasmas generate two counter-propagating collisionless shock waves \citep{ForslundPRL1970,SilvaApJ2003,Spitkovsky2008a,RyutovlPPCF2018}. In the collisional case, the shocks are launched when the two fluids make contact. In the collisionless case, the two plasmas start interpenetrating as the long mean-free-path prevent them from ``bumping'' into each other. As a counter-streaming plasma system, the interpenetrating region quickly turns unstable. The instability grows, saturates, and creates a localized turbulence which stops the incoming flow, initiating the density build-up in the overlapping region \citep{BretPoP2013,BretPoP2014,DieckmannJPP2017}.

Various kind of instabilities do grow in the overlapping region \citep{BretPoPReview}. But the fastest growing one takes the lead and eventually defines the ensuing turbulence. When the system is such that the filamentation, or Weibel, instability grows faster, magnetic filaments are generated \citep{Medvedev1999,wiersma04,lyubarsky06,kato2007,PRLLemoine2019}. In a pair plasma\footnote{To our knowledge, there is no systematic study of the hierarchy of unstable modes for two magnetized colliding ion/electron plasma shells. See \cite{lyubarsky06,Yalinewich2010,Shaisultanov2012} for works contemplating the un-magnetized case.}, the conditions required for the Weibel instability to lead the linear phase have been studied in \cite{BretPoP2016} for a flow-aligned field, and in \cite{BretPoP2017} for an oblique field. A mildly relativistic flow is required. Also, accounting for an oblique field supposes the Larmor radius of the particles is large compared to the dimensions of the system. Since the field modifies the hierarchy of unstable modes, it can prompt another mode than Weibel to lead the linear phase. In such cases, studies found so far that a shock still forms \citep{BretJPP2017,DieckmannJPP2017,DieckmannMNRAS2018}, mediated by the growth of the non-Weibel leading instability, like two-stream for example.

Since the blocking of the flow entering the filaments is key for the shock formation, it is interesting to study under which conditions a test particle will be stopped in these magnetic filaments. Recently, a toy model of the process successfully reproduced the criteria for shock formation in the case of pair plasmas \citep{BretPoP2015}. When implemented accounting for a flow-aligned magnetic field \citep{BretJPP2016}, the same kind of model could predict how too strong a field can deeply affect the shock formed \citep{BretJPP2017,BretJPP2018}. In view of the many settings involving oblique magnetic fields, we extend here the previous model to the case of an oblique field.

\begin{figure}
\includegraphics[width=\textwidth]{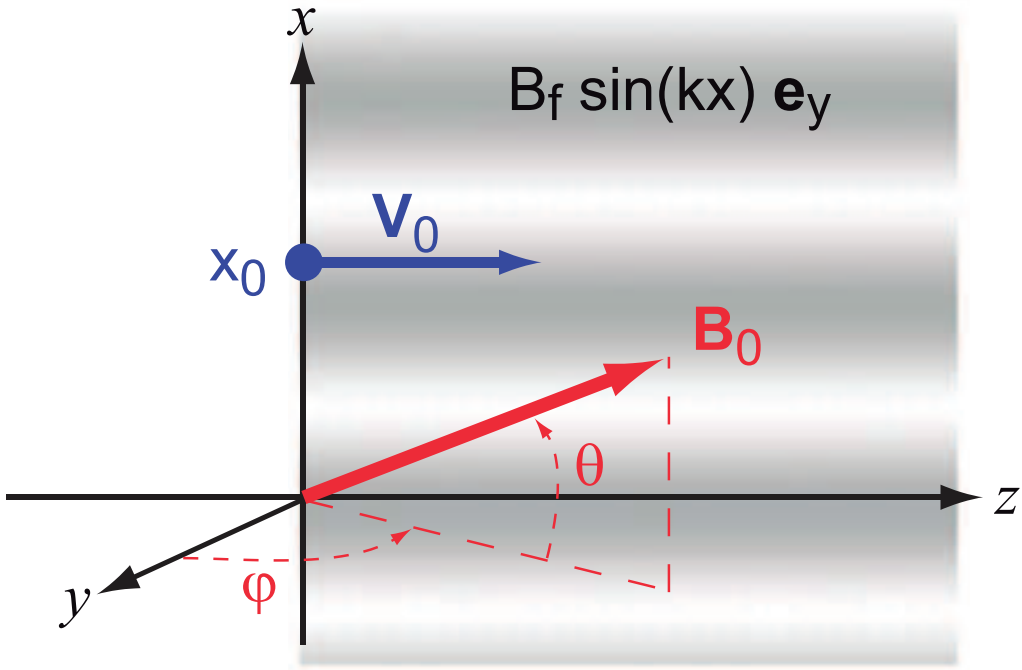}
\caption{Setup considered. We can consider $\varphi=\pi/2$ since the direction of the flow, the field, and the $\mathbf{k}$ of the fastest growing Weibel mode, are coplanar for the fastest growing Weibel modes \citep{BretPoP2014a,Stockem2016}.}
\label{setup}
\end{figure}

The system considered is sketched on Figure \ref{setup}. The half space $z\geq 0$ is filled with the magnetic filaments $\mathbf{B}_f=B_f\sin ( kx) ~\mathbf{e}_y$. In principle, we should consider any possible orientation for $\mathbf{B}_0$, thus having to consider the angles $\theta$ and $\varphi$. Yet, previous works in 3D geometry found that the fastest growing Weibel modes are found with a wave vector coplanar with the field $\mathbf{B}_0$ and the direction of the flow \citep{BretPoP2014a,Stockem2016}. Here, the initial flow giving rise to Weibel is along $z$ and we set up the axis so that $\mathbf{k}$ is along $x$. We can therefore consider $\varphi=\pi/2$ so that $\mathbf{B}_0=B_0(\sin\theta,0,\cos\theta)$.

A test particle is injected at $(x_0,0,0)$ with velocity $\mathbf{v}_0=(0,0,v_0)$ and Lorentz factor $\gamma_0=(1-v_0^2/c^2)^{-1/2}$, mimicking a particle of the flow entering the filamented overlapping region. Our goal is to determine under which conditions the particle streams through the magnetic filaments to $z=+\infty$, or is trapped inside. As explained in previous articles \citep{BretPoP2015,BretJPP2016}, in the present toy model, this dichotomy comes down to determining whether test particles stream through the filaments, or bounce back to the region $z<0$. The reason for this is that in a more realistic setting, particles reaching the filaments and turning back will likely be trapped in the turbulent region between the upstream and the downstream. This problem is clearly related to the one of the shock formation, since the density build-up leading to the shock requires the incoming flow to be trapped in the filaments.

\section{Equations of motion}
Since the Lorentz factor $\gamma_0$ is a constant of the motion, the equation of motion for our test particle reads,
\begin{equation}\label{eq:motion}
  m\gamma_0 \ddot{\mathbf{x}} = q\frac{\dot{\mathbf{x}}}{c}\times (\mathbf{B}_f + \mathbf{B}_0).
\end{equation}
Explaining each component, we find for $z>0$,
\begin{eqnarray}
  \ddot{x} &=&  \frac{q B_f}{\gamma_0 m c}  \left[ \frac{B_0}{B_f}  \dot{y} \cos\theta  - \dot{z} \sin kx
   \right],  \label{eq:motionx} \\
  \ddot{y} &=&    \frac{q B_f}{\gamma_0 m c}\frac{B_0}{B_f}\left[
          \dot{z} \sin\theta   - \dot{x}\cos\theta
  \right] ,  \label{eq:motiony} \\
  \ddot{z} &=&     \frac{q B_f}{\gamma_0 m c} \left[-\frac{B_0}{B_f} \dot{y} \sin\theta  + \dot{x} \sin kx   \right], \label{eq:motionz}
\end{eqnarray}
while $B_f=0$ for $z<0$. We can now define the following dimensionless variables,
\begin{equation}\label{eq:change}
  \mathbf{X}=k\mathbf{x},~~\alpha = \frac{B_0}{B_f}, ~~ \tau = t\omega_{B_f},~~\mathrm{with}  ~~ \omega_{B_f} = \frac{qB_f}{\gamma_0 mc}.
\end{equation}
With these variables, the system (\ref{eq:motionx}-\ref{eq:motionz}) reads,
\begin{eqnarray}
\ddot{X} =&  -\dot{Z} \mathcal{H}(Z) \sin X + & \alpha   \dot{Y} \cos\theta   ,  \label{eq:dimless1}\nonumber \\
\ddot{Y} =&                    & \alpha  ( \dot{Z} \sin\theta  - \dot{X} \cos\theta ),   \label{eq:dimless2}\\
\ddot{Z} =&  \dot{X} \mathcal{H}(Z) \sin X -  & \alpha   \dot{Y} \sin\theta   , \label{eq:dimless3}\nonumber
\end{eqnarray}
where $\mathcal{H}$ is the Heaviside step function $\mathcal{H}(x)=0$ for $x<0$ and $\mathcal{H}(x)=1$ for $x \geq 0$. The  initial conditions are,
\begin{eqnarray}\label{eq:ini}
\mathbf{X}(\tau=0)         &=& \left( X_0,0,0 \right)             ~~ \mathrm{with} ~~ X_0       \equiv  kx_0,  \nonumber \\
\dot{\mathbf{X}}(\tau=0)  &=& \left(0,0,\dot{Z}_0 \right) ~~ \mathrm{with} ~~ \dot{Z}_0 \equiv \frac{kv_0}{\omega_{B_f}}.
\end{eqnarray}

\section{Constants of the motion and chaotic behavior}\label{sec:chaos}
The total field in the region $z>0$ reads $\mathbf{B}=\left( B_0 \sin\theta, B_f\sin ( kx), B_0 \cos\theta \right)$. It can be written as $\mathbf{B}=\nabla \times \mathbf{A}$, with the vector potential in the Coulomb gauge,
\begin{equation}\label{eq:A}
\mathbf{A} = \left(
\begin{array}{c}
  0 \\
  B_0 x \cos\theta \\
  B_0 y \sin\theta + B_f\frac{\cos (kx)}{k}
\end{array}
  \right).
\end{equation}
The canonical momentum then reads \citep{jackson1998},
\begin{equation}\label{eq:P}
\mathbf{P} = \mathbf{p} + \frac{q}{c}\mathbf{A} = \left(
\begin{array}{c}
  p_x \\
  p_y + \frac{q}{c} B_0 x \cos\theta \\
  p_z +  \frac{q}{c}\left(  B_0 y \sin\theta + B_f\frac{\cos (kx)}{k} \right)
\end{array}
  \right),
\end{equation}
where $\mathbf{p}=\gamma_0 m \mathbf{v}$. Since it does not explicitly depend on $z$, the $z$ component is a constant of the motion. It can equally be obtained time-integrating Eq. (\ref{eq:motionz}). Note that for $\theta=0$, the $y$ dependence vanishes so that the $y$ component of the canonical momentum is also a constant of the motion \citep{BretJPP2016}.

We can derive another constant of the motion from Eq. (\ref{eq:motiony}). Time-integrating it and remembering $\dot{y}(t=0)=z(t=0)=0$, we find,
\begin{equation}
 \dot{y} =    \frac{q B_0}{\gamma_0 m c}\left[
          z \sin\theta   - (x-x_0)\cos\theta
  \right].
\end{equation}
Since $x_0$ is obviously a constant, we can express it in terms of the other variables and obtain an invariant, that is
\begin{eqnarray}\label{eq:x0}
 x_0 &=&  x - z \tan\theta + \frac{ c}{q B_0 \cos\theta}\gamma_0 m\dot{y} \nonumber \\
     &=& x - z \tan\theta + \frac{ c}{q B_0 \cos\theta}\left(P_y -\frac{q}{c}A_y\right).
\end{eqnarray}
Finally, the Hamiltonian,
\begin{equation}\label{eq:H}
 \mathcal{ H} = c \sqrt{c^2 m^2 + \left( \mathbf{P} - \frac{q}{c}\mathbf{A} \right)^2}
\end{equation}
is also a constant of the motion. Replacing $A_y$ in Eq. (\ref{eq:x0}) by its expression from (\ref{eq:A}), we therefore have the following constants of the motion,
\begin{eqnarray}
  \mathcal{H} \equiv C_1 &=& c \sqrt{c^2 m^2 + \left( \mathbf{P} - \frac{q}{c}\mathbf{A} \right)^2}, \\
  C_2                    &=& P_z, \\
  x_0 \equiv C_3         &=& \frac{ c}{q B_0 \cos\theta}P_y - z \tan\theta .
\end{eqnarray}

According to Liouville's theorem on integrable systems, a $n$-dimensional Hamiltonian system is \emph{integrable} if it has $n$ constants of motion $C_j(x_i,P_i,t)_{j \in [1\ldots n]}$ in involution \citep{ott2002chaos,lichtenberg2013Chaos,ChenJGR1986}, that is,
\begin{equation}\label{eq:Poisson}
\{C_j,C_k\} = \sum_{i=1}^3 \left(\frac{\partial C_j}{\partial x_i}\frac{\partial C_k}{\partial P_i} - \frac{\partial C_k}{\partial x_i}\frac{\partial C_j}{\partial P_i}\right) = 0, \forall (j,k),
\end{equation}
where  $\{f,g\}$ is the Poisson bracket of $f$ and $g$. It is easily checked that  $C_{2,3}$ being constants of the motion, $\{\mathcal{ H},C_2\}=\{\mathcal{ H},C_3\}=0$. However,
\begin{equation}\label{eq:C2C3}
\{C_2,C_3\} = -\tan\theta.
\end{equation}
As a result, the system is integrable only for $\theta=0$ \citep{BretJPP2016}. Otherwise, it is chaotic, as will be checked numerically in the following sections.

\section{Reduction of the number of free parameters}
The free parameters of the system (\ref{eq:motionx}-\ref{eq:motionz}) with initial conditions (\ref{eq:ini}) are $(X_0,\dot{Z}_0,\alpha,\theta)$. In order to deal with a more tractable phase space parameter, we now reduce its dimension accounting for the physical context of the problem.

Consider the magnetic filaments generated by the growth of the filamentation instability triggered by the counter-streaming of 2 cold (thermal spread $\Delta v \ll v$) symmetric pair plasmas. Both plasma shells have identical density $n$ in the lab frame, and initial velocities $\pm v \mathbf{e}_z$. We denote $\beta = v/c$. The Lorentz factor $\gamma_0$ previously defined equally reads $\gamma_0=(1-\beta^2)^{-1/2}$ since the test particles entering the magnetic filaments belong to the same plasma shells.

The wave vector $\mathbf{k}$ defining the magnetic filaments is also the wave vector of the fastest growing filamentation modes. We can then set \citep{BretPoP2013},
\begin{equation}\label{eq:km}
k = \frac{\omega_p}{c\sqrt{\gamma_0}},
\end{equation}
where $\omega_p^2=4\pi n q^2/m$, so that,
\begin{equation}\label{eq:Z0reduit}
\dot{Z}_0 = \frac{\beta}{\sqrt{\gamma_0}} \frac{\omega_p}{\omega_{B_f} }
          = \frac{\beta}{\sqrt{\gamma_0}} \frac{\omega_{B_0}}{\omega_{B_f}}        \frac{\omega_p}{\omega_{B_0} }
          = \frac{\beta}{\sqrt{\gamma_0}} \alpha                            \frac{\omega_p}{\omega_{B_0} },
\end{equation}
where,
\begin{equation}
\omega_{B_0}=\frac{qB_0}{\gamma_0 mc}.
\end{equation}
The peak field $B_f$ in the filaments can be estimated from the growth rate $\delta$ of the instability, considering $\omega_{B_f} \sim \delta$ \citep{davidsonPIC1972}. It turns out that over the domain $\delta \gg \omega_{B_0}$, the growth rate $\delta$ depends weakly on $\theta$ and can be well approximated by \citep{StockemApJ2006,BretPoP2014a},
\begin{equation}
\omega_{B_f} \sim \delta = \omega_p\sqrt{\frac{2 \beta^2}{\gamma_0} - \left( \frac{\omega_{B_0}}{\omega_p}\right)^2},
\end{equation}
so that,
\begin{equation}
\omega_{B_f} = \frac{\omega_{B_0}}{\alpha} =  \omega_p\sqrt{\frac{2 \beta^2}{\gamma_0} - \left( \frac{\omega_{B_0}}{\omega_p}\right)^2}.
\end{equation}
This expression allows one to express $\omega_{B_0}/\omega_p$ as,
\begin{equation}
 \frac{\omega_{B_0}}{\omega_p} = \sqrt{\frac{2}{\gamma_0}} \frac{\alpha \beta}{\sqrt{1+\alpha^2  }},
\end{equation}
so that Eq. (\ref{eq:Z0reduit}) eventually reads,
\begin{equation}\label{eq:Z0reduitOK}
  \dot{Z}_0 = \sqrt{\frac{1+\alpha^2 }{2}} .
\end{equation}
The parameters phase space is thus reduced to 3 dimensions, $(X_0,\alpha,\theta)$.

\section{Numerical exploration}
It was previously found that for $\theta=0$, all particles stream through the filaments, no matter their initial position and velocity, if $\alpha > 1/2$ \citep{BretJPP2016}. Clearly for $\theta=\pi/2$, no particle can stream to $z=+\infty$. As we shall see, the $\theta$-dependent threshold value of $\alpha$ beyond which all particles go to $\infty$ is simply $\propto 1/\cos\theta$.

The system (\ref{eq:dimless1}-\ref{eq:dimless3}) is solved using the \emph{Mathematica} ``NDSolve'' function. The equations are invariant under the change $X \rightarrow X+2 \pi$, so that we can restrict the investigation to $X_0 \in [-\pi, \pi]$. Unless $\theta = 0$, there are no other trivial symmetries. In particular, the transformation $\theta \rightarrow -\theta$ does \emph{not} leave the system invariant. We shall detail the case $\theta \in [0,\pi/2]$ and only give the results, very similar though not identical, for $\theta \in [-\pi/2,0]$.

\begin{figure}
\begin{center}
 \includegraphics[width=0.48\textwidth]{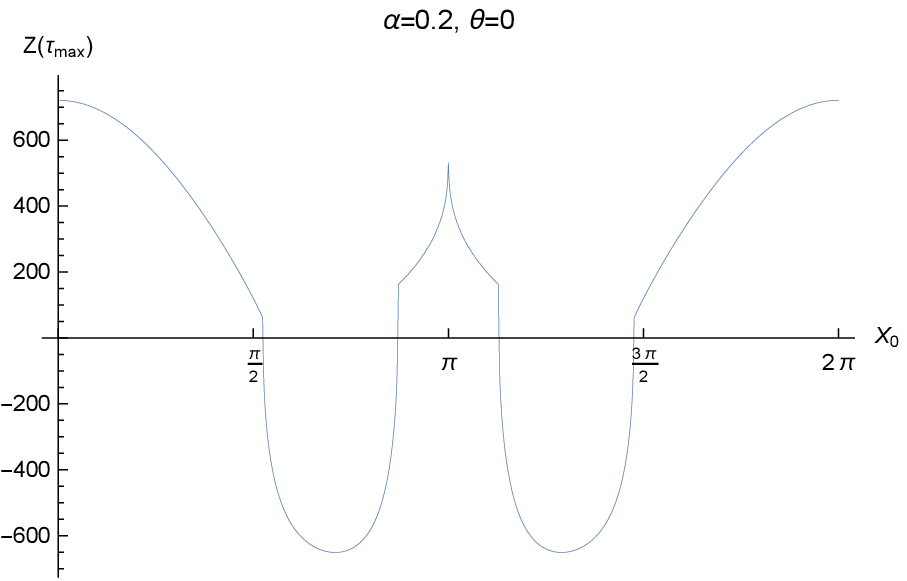} \includegraphics[width=0.48\textwidth]{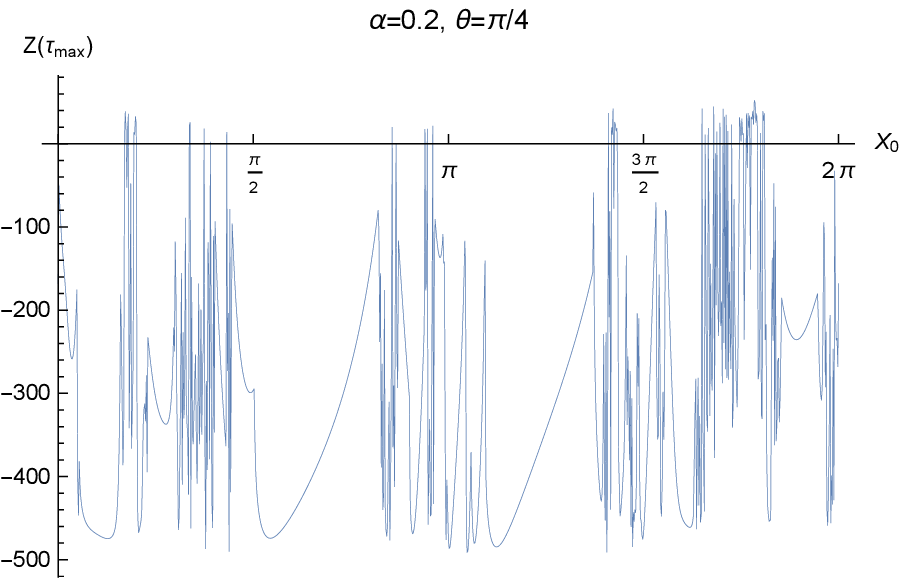}
\end{center}
\caption{Value of $Z(\tau_{max})$ in terms of $X_0$ for $\tau_{max}=10^3$ and two values of $\theta$.}\label{fig2}
\end{figure}

The numerical exploration is conducted solving the equations and looking for the value of $Z$ at large time $\tau=\tau_{max}$. Figure \ref{fig2} shows the value of $Z(\tau_{max})$ in terms of $X_0 \in [0,2\pi]$ for the specified values of $(\alpha,\theta)$ and $\tau_{max}=10^3$. Similar results have been obtained for larger values of $\tau$ like $\tau=10^4$, or even smaller ones, like $\tau=500$. For $\theta=\pi/4$, save a few exceptions for some values of $X_0$, all particles have bounced back to the $z<0$ region. As explained in  \cite{BretPoP2015,BretJPP2016}, this means that in a more realistic setting, they would likely be trapped in the filaments.

\begin{figure}
\begin{center}
 \includegraphics[width=0.32\textwidth]{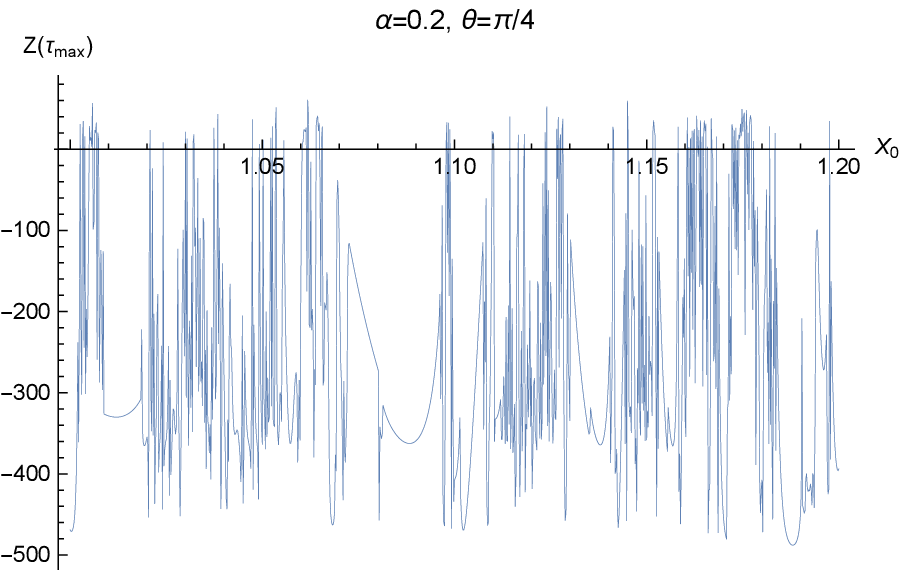}\includegraphics[width=0.32\textwidth]{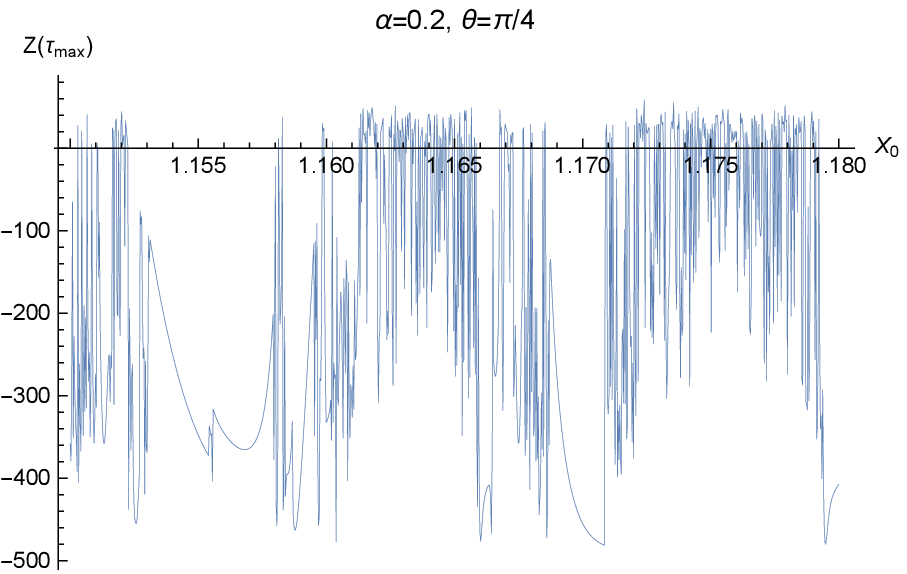}\includegraphics[width=0.32\textwidth]{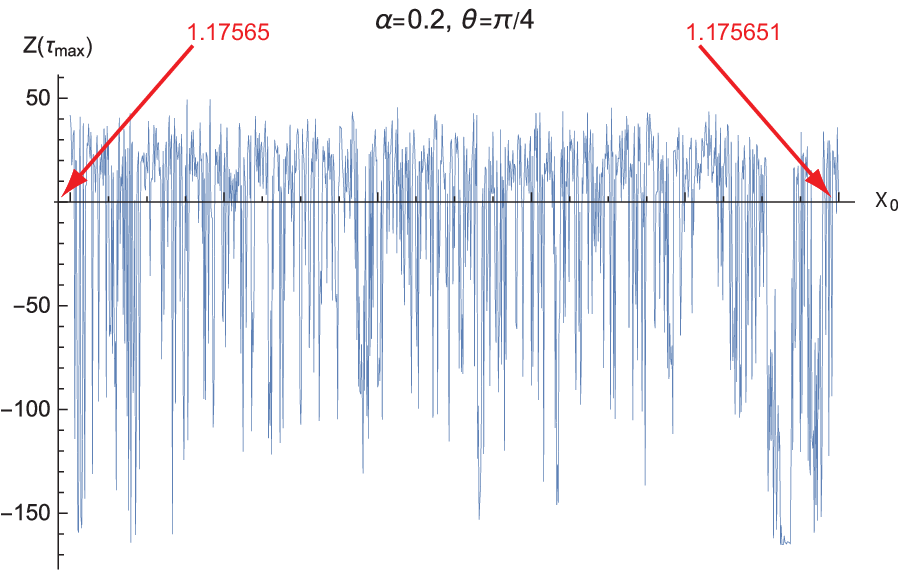}
\end{center}
\caption{Value of $Z(\tau_{max})$ for $\theta=\pi/4$ and increasingly small $X_0$ intervals.}\label{fig3}
\end{figure}

As evidenced by Figure \ref{fig2}-left, the function $Z(\tau_{max})$ is smooth for $\theta=0$. Yet, for $\theta=\pi/4$ (right), the result features regions where $Z(\tau_{max})$ varies strongly with $X_0$. In order to identify chaos, Figure \ref{fig3} present a series of successive zooms on Fig. \ref{fig2}-right, where the function $Z(\tau_{max})$ is plotted over an increasingly small $X_0$ interval inside $X_0 \in [\pi/4,\pi/2]$.

As expected from the analysis conducted in Sec. \ref{sec:chaos}, the system is chaotic. Note that chaotic trajectories in magnetic field lines have already been identified in literature \citep{ChenJGR1986,Buchner1989,PoPAbhay2010,ChaosCambon2014}.

Figure \ref{fig2} has been plotted solving the system for $N+1$ particles shot from $X_0(j) = j 2\pi/N$ with $j=0\ldots N$. We denote $Z_j(\tau_{max})$ the value of $Z$ reached by the $j^{\mathrm{th}}$ particle at $\tau=\tau_{max}$. Then we define the following function,
\begin{equation}\label{eq:phi}
  \phi (\alpha,\theta) = \frac{1}{N+1}\sum_{j=0}^{N} \mathcal{H}\left[  -Z_j(\tau_{max})  \right],
\end{equation}
where $\mathcal{H}$ is again the Heaviside function.

\begin{figure}
\begin{center}
 \includegraphics[width=0.48\textwidth]{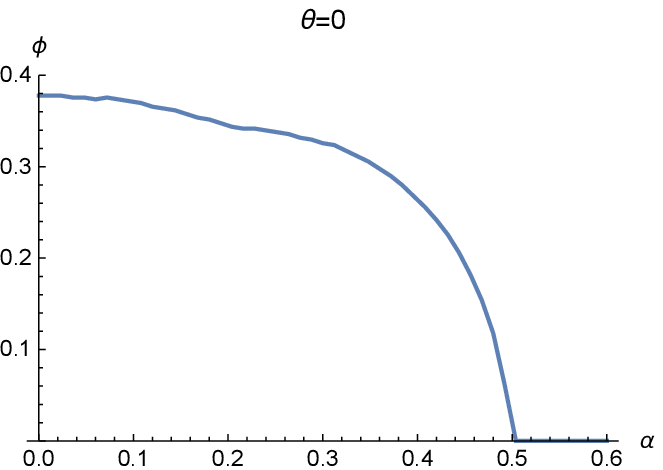}\includegraphics[width=0.48\textwidth]{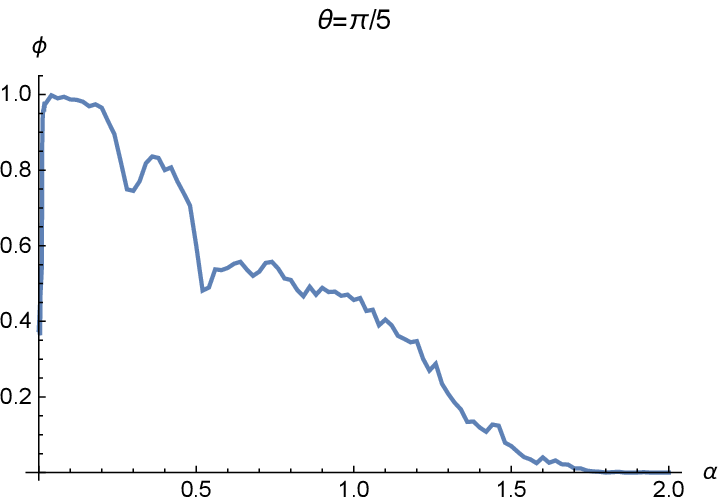}
\end{center}
\caption{Plot on the function $\phi$ defined by Eq. (\ref{eq:phi}), in terms of $\alpha$ and for $\theta=0$ and $\pi/5$.}\label{fig4}
\end{figure}

The function $\phi$ represents therefore the fraction of particles that bounced back against the magnetic filaments. Figure \ref{fig4}-left plots it in terms of $\alpha$ for $\theta=0$. For $\alpha=0$, that is $B_0=0$, about 40\% of the particles bounce back, i.e, are trapped in the filaments. As $\alpha$ is increased, the field $\mathbf{B}_0$ guides the test particles more and more efficiently until $\alpha \sim 0.5$ where all the particles stream through the filaments. In turn, Figure \ref{fig4}-right displays the case $\theta=\pi/5$. Being oblique, the field $\mathbf{B}_0$ is less efficient to guide the particles through the filaments, and more efficient to trap them inside. As a result, it takes a higher value of $B_0$, that is, $\alpha = 1.4$, to reach $\phi=0$.

\begin{figure}
\begin{center}
 \includegraphics[width=\textwidth]{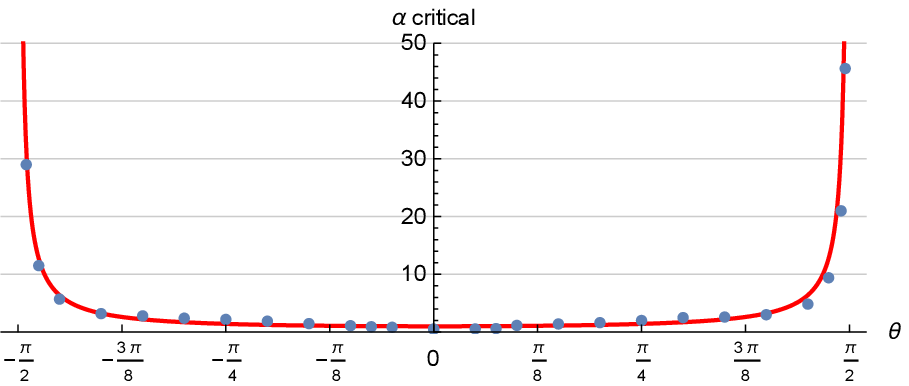}\\ \includegraphics[width=\textwidth]{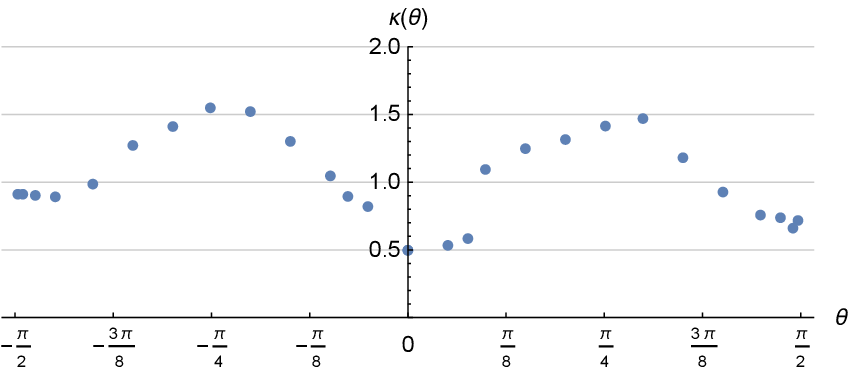}
\end{center}
\caption{\emph{Top}: Numerical computation of $\alpha_c(\theta)$ (blue dots) compared to $1/\cos\theta$ (red line). \emph{Bottom}: Values of the $\kappa (\theta)$ function entering the expression of $\alpha_c$ in Eq. (\ref{eq:aplha_c}), in terms of $\theta$.}\label{fig5}
\end{figure}

Similar numerical calculations have been conducted for various values of $\theta \in [0,\pi/2]$. We finally define $\alpha_c(\theta)$ such as,
\begin{equation}\label{eq:aplha_cDef}
  \phi(\alpha_c)=0.
\end{equation}
$\alpha_c(\theta)$ is therefore the threshold value of $\alpha$ beyond which all particles bounce back against the filamented region, i.e, $\phi(\alpha\geq\alpha_c)=0$. It is plotted on Figure \ref{fig5} and can be well approximated by,
\begin{equation}\label{eq:aplha_c}
  \alpha_c = \kappa (\theta) \frac{1}{\cos\theta},
\end{equation}
where $\kappa (\theta)$ is of order unity (see Fig. \ref{fig5}-bottom).

\begin{figure}
\begin{center}
 \includegraphics[width=\textwidth]{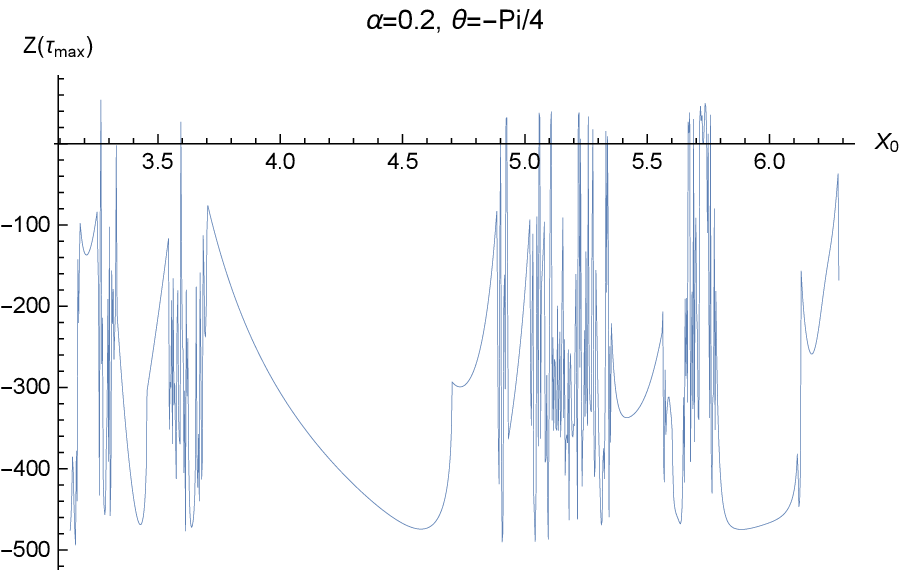}
\end{center}
\caption{Same as Fig. \ref{fig2}-right, but for $\theta=-\pi/4$.}\label{fig6}
\end{figure}

\subsection{Case $\theta < 0$}
As already noticed, there is no invariance by the change $\theta \rightarrow -\theta$. Figure \ref{fig6} plots the counterpart of Fig. \ref{fig2}-right, but for $\theta=-\pi/4$. Though quite similar, the results are not identical. The numerical analysis detailed above for $\theta \in [0,\pi/2]$ has been conducted for $\theta \in [-\pi/2,0]$. The function $\alpha_c(\theta<0)$ again adjusts very well to the right-hand-side of Eq. (\ref{eq:aplha_c}), with the values of $\kappa (\theta)$ plotted on Fig. \ref{fig5}-bottom.

\section{Conclusion}
A model previously developed to study test particles  trapping in magnetic filaments has been extended to the case of an oblique external magnetic field. The result makes perfect physical sense: up to a constant $\kappa$ of order unity, only the component of the field parallel to the filaments is relevant. The criteria obtained in \cite{BretJPP2016} for a flow-aligned field, namely that particles are trapped if $\alpha > 1/2$, now reads,
\begin{equation}
  \alpha  > \frac{\kappa(\theta)}{\cos\theta},
\end{equation}
where $\kappa(\theta)$ is of order unity. As a consequence, a parallel field affects the shock more than an oblique one. Kinetic effects triggered by a parallel field can significantly modify  the shock structure while a perpendicular field will rather ``help'' the shock formation. This is in agreement with theoretical works which found a parallel field can divide by 2 the density jump expected from the MHD Rankine-Hugoniot conditions \citep{BretJPP2017,BretJPP2018}, while the departure from MHD is less pronounced in the perpendicular case \citep{BretPoP2019}.

\section{Acknowledgments}
A.B. acknowledges support by grants ENE2016-75703-R from the Spanish
Ministerio de Educaci\'{o}n and SBPLY/17/180501/000264 from the Junta de
Comunidades de Castilla-La Mancha. Thanks are due to Ioannis Kourakis and Didier B\'{e}nisti for enriching discussions.


\begin{thebibliography}{36}
\expandafter\ifx\csname natexlab\endcsname\relax\def\natexlab#1{#1}\fi

\bibitem[Bret(2014)]{BretPoP2014a}
{\sc Bret, A.} 2014 Robustness of the filamentation instability in arbitrarily
  oriented magnetic field: Full three dimensional calculation. {\em Physics of
  Plasmas\/} {\bf 21}~(2), 022106.

\bibitem[Bret(2015)]{BretPoP2015}
{\sc Bret, A.} 2015 Particles trajectories in magnetic filaments. {\em Physics
  of Plasmas\/} {\bf 22}, 072116.

\bibitem[Bret(2016{\natexlab{{\em a\/}}})]{BretPoP2016}
{\sc Bret, A.} 2016{\natexlab{{\em a\/}}} Hierarchy of instabilities for two
  counter-streaming magnetized pair beams. {\em Physics of Plasmas\/} {\bf 23},
  062122.

\bibitem[Bret(2016{\natexlab{{\em b\/}}})]{BretJPP2016}
{\sc Bret, A.} 2016{\natexlab{{\em b\/}}} Particles trajectories in weibel
  magnetic filaments with a flow-aligned magnetic field. {\em Journal of Plasma
  Physics\/} {\bf 82}, 905820403.

\bibitem[Bret \& Dieckmann(2017)]{BretPoP2017}
{\sc Bret, A. \& Dieckmann, M.~E.} 2017 Hierarchy of instabilities for two
  counter-streaming magnetized pair beams: Influence of field obliquity. {\em
  Physics of Plasmas\/} {\bf 24}~(6), 062105.

\bibitem[Bret {\em et~al.\/}(2010)Bret, Gremillet \& Dieckmann]{BretPoPReview}
{\sc Bret, A., Gremillet, L. \& Dieckmann, M.~E.} 2010 Multidimensional
  electron beam-plasma instabilities in the relativistic regime. {\em Phys.
  Plasmas\/} {\bf 17}, 120501.

\bibitem[{Bret} \& {Narayan}(2018)]{BretJPP2018}
{\sc {Bret}, Antoine \& {Narayan}, Ramesh} 2018 {Density jump as a function of
  magnetic field strength for parallel collisionless shocks in pair plasmas}.
  {\em Journal of Plasma Physics\/} {\bf 84}, 905840604.

\bibitem[{Bret} \& {Narayan}(2019)]{BretPoP2019}
{\sc {Bret}, A. \& {Narayan}, R.} 2019 {Density jump as a function of magnetic
  field for collisionless shocks in pair plasmas: The perpendicular case}. {\em
  Physics of Plasmas\/} {\bf 26}, 062108.

\bibitem[{Bret} {\em et~al.\/}(2017){Bret}, {Pe'er}, {Sironi}, {S\c{a}dowski}
  \& {Narayan}]{BretJPP2017}
{\sc {Bret}, Antoine, {Pe'er}, Asaf, {Sironi}, Lorenzo, {S\c{a}dowski},
  Aleksander \& {Narayan}, Ramesh} 2017 {Kinetic inhibition of
  magnetohydrodynamics shocks in the vicinity of a parallel magnetic field}.
  {\em Journal of Plasma Physics\/} {\bf 83}, 715830201.

\bibitem[{Bret} {\em et~al.\/}(2013){Bret}, {Stockem}, {Fiuza}, {Ruyer},
  {Gremillet}, {Narayan} \& {Silva}]{BretPoP2013}
{\sc {Bret}, A., {Stockem}, A., {Fiuza}, F., {Ruyer}, C., {Gremillet}, L.,
  {Narayan}, R. \& {Silva}, L.~O.} 2013 {Collisionless shock formation,
  spontaneous electromagnetic fluctuations, and streaming instabilities}. {\em
  Physics of Plasmas\/} {\bf 20}~(4), 042102.

\bibitem[Bret {\em et~al.\/}(2014)Bret, Stockem, Narayan \& Silva]{BretPoP2014}
{\sc Bret, A., Stockem, A., Narayan, R. \& Silva, L.~O.} 2014 Collisionless
  weibel shocks: Full formation mechanism and timing. {\em Physics of
  Plasmas\/} {\bf 21}~(7), 072301.

\bibitem[B\"{u}chner \& Zelenyi(1989)]{Buchner1989}
{\sc B\"{u}chner, J\"{o}rg \& Zelenyi, Lev~M.} 1989 Regular and chaotic charged
  particle motion in magnetotaillike field reversals: 1. basic theory of
  trapped motion. {\em Journal of Geophysical Research: Space Physics\/} {\bf
  94}~(A9), 11821--11842.

\bibitem[Cambon {\em et~al.\/}(2014)Cambon, Leoncini, Vittot, Dumont \&
  Garbet]{ChaosCambon2014}
{\sc Cambon, Benjamin, Leoncini, Xavier, Vittot, Michel, Dumont, Rémi \&
  Garbet, Xavier} 2014 Chaotic motion of charged particles in toroidal magnetic
  configurations. {\em Chaos\/} {\bf 24}~(3), 033101.

\bibitem[{Chen} \& {Palmadesso}(1986)]{ChenJGR1986}
{\sc {Chen}, J. \& {Palmadesso}, P.~J.} 1986 {Chaos and nonlinear dynamics of
  single-particle orbits in a magnetotaillike magnetic field}. {\em Journal of
  Geophysical Research\/} {\bf 91}, 1499--1508.

\bibitem[Davidson {\em et~al.\/}(1972)Davidson, Hammer, Haber \&
  Wagner]{davidsonPIC1972}
{\sc Davidson, Ronald~C., Hammer, David~A., Haber, Irving \& Wagner, Carl~E.}
  1972 Nonlinear development of electromagnetic instabilities in anisotropic
  plasmas. {\em Phys. Fluids\/} {\bf 15}, 317.

\bibitem[{Dieckmann} \& {Bret}(2017)]{DieckmannJPP2017}
{\sc {Dieckmann}, M.~E. \& {Bret}, A.} 2017 {Simulation study of the formation
  of a non-relativistic pair shock}. {\em Journal of Plasma Physics\/} {\bf
  83}~(1), 905830104.

\bibitem[Dieckmann \& Bret(2018)]{DieckmannMNRAS2018}
{\sc Dieckmann, M.~E. \& Bret, A.} 2018 Electrostatic and magnetic
  instabilities in the transition layer of a collisionless weakly relativistic
  pair shock. {\em Monthly Notices of the Royal Astronomical Society\/} {\bf
  473}~(1), 198--209.

\bibitem[Forslund \& Shonk(1970)]{ForslundPRL1970}
{\sc Forslund, D.~W. \& Shonk, C.~R.} 1970 Formation and structure of
  electrostatic collisionless shocks. {\em Phys. Rev. Lett.\/} {\bf 25},
  1699--1702.

\bibitem[Jackson(1998)]{jackson1998}
{\sc Jackson, J.D.} 1998 {\em Classical Electrodynamics\/}. Wiley.

\bibitem[Kato(2007)]{kato2007}
{\sc Kato, Tsunehiko~N.} 2007 Relativistic collisionless shocks in unmagnetized
  electron-positron plasmas. {\em The Astrophysical Journal\/} {\bf 668}~(2),
  974.

\bibitem[Lemoine {\em et~al.\/}(2019)Lemoine, Gremillet, Pelletier \&
  Vanthieghem]{PRLLemoine2019}
{\sc Lemoine, Martin, Gremillet, Laurent, Pelletier, Guy \& Vanthieghem, Arno}
  2019 Physics of weibel-mediated relativistic collisionless shocks. {\em Phys.
  Rev. Lett.\/} {\bf 123}, 035101.

\bibitem[Lichtenberg \& Lieberman(2013)]{lichtenberg2013Chaos}
{\sc Lichtenberg, A.J. \& Lieberman, M.A.} 2013 {\em Regular and Chaotic
  Dynamics\/}. Springer New York.

\bibitem[Lyubarsky \& Eichler(2006)]{lyubarsky06}
{\sc Lyubarsky, Y. \& Eichler, D.} 2006 Are {Gamma}-ray bursts mediated by the
  weibel instability? {\em Astrophys. J.\/} {\bf 647}, 1250.

\bibitem[Medvedev \& Loeb(1999)]{Medvedev1999}
{\sc Medvedev, M.~V. \& Loeb, A.} 1999 Generation of magnetic fields in the
  relativistic shock of gamma-ray burst sources. {\em Astrophys. J.\/} {\bf
  526}, 697.

\bibitem[Novo {\em et~al.\/}(2016)Novo, Bret \& Sinha]{Stockem2016}
{\sc Novo, A~Stockem, Bret, A \& Sinha, U} 2016 Shock formation in magnetised
  electron{\textendash}positron plasmas: mechanism and timing. {\em New Journal
  of Physics\/} {\bf 18}~(10), 105002.

\bibitem[Ott(2002)]{ott2002chaos}
{\sc Ott, E.} 2002 {\em Chaos in Dynamical Systems\/}. Cambridge University
  Press.

\bibitem[Ram \& Dasgupta(2010)]{PoPAbhay2010}
{\sc Ram, Abhay~K. \& Dasgupta, Brahmananda} 2010 Dynamics of charged particles
  in spatially chaotic magnetic fields. {\em Physics of Plasmas\/} {\bf
  17}~(12), 122104.

\bibitem[Ryutov(2018)]{RyutovlPPCF2018}
{\sc Ryutov, D~D} 2018 Collisional and collisionless shocks. {\em Plasma
  Physics and Controlled Fusion\/} {\bf 61}~(1), 014034.

\bibitem[Sagdeev \& Kennel(1991)]{Sagdeev_Kennel_1991}
{\sc Sagdeev, R.Z. \& Kennel, C.F.} 1991 Collisionless shock waves. {\em
  Scientific American; (United States)\/} {\bf 264:4}.

\bibitem[Shaisultanov {\em et~al.\/}(2012)Shaisultanov, Lyubarsky \&
  Eichler]{Shaisultanov2012}
{\sc Shaisultanov, R., Lyubarsky, Y. \& Eichler, D.} 2012 Stream instabilities
  in relativistically hot plasma. {\em The Astrophysical Journal\/} {\bf 744},
  182.

\bibitem[Silva {\em et~al.\/}(2003)Silva, Fonseca, Tonge, Dawson, Mori \&
  Medvedev]{SilvaApJ2003}
{\sc Silva, L.~O., Fonseca, R.~A., Tonge, J.~W., Dawson, J.~M., Mori, W.~B. \&
  Medvedev, M.~V.} 2003 Interpenetrating plasma shells: Near-equipartition
  magnetic field generation and nonthermal particle acceleration. {\em
  Astrophys. J.\/} {\bf 596}, L121--L124.

\bibitem[Spitkovsky(2008)]{Spitkovsky2008a}
{\sc Spitkovsky, Anatoly} 2008 Particle acceleration in relativistic
  collisionless shocks: Fermi process at last? {\em Astrophys. J. Lett.\/} {\bf
  682}, L5--L8.

\bibitem[Stockem {\em et~al.\/}(2006)Stockem, Lerche \&
  Schlickeiser]{StockemApJ2006}
{\sc Stockem, A., Lerche, I. \& Schlickeiser, R.} 2006 On the physical
  realization of two-dimensional turbulence fields in magnetized interplanetary
  plasmas. {\em The Astrophysical Journal\/} {\bf 651}~(1), 584.

\bibitem[Wiersma \& Achterberg(2004)]{wiersma04}
{\sc Wiersma, J. \& Achterberg, A.} 2004 Magnetic field generation in
  relativistic shocks. an early end of the exponential weibel instability in
  electron-proton plasmas. {\em Astron. Astrophys.\/} {\bf 428}, 365--371.

\bibitem[Yalinewich \& Gedalin(2010)]{Yalinewich2010}
{\sc Yalinewich, A. \& Gedalin, M.} 2010 Instabilities of relativistic
  counterstreaming proton beams in the presence of a thermal electron
  background. {\em Phys. Plasmas\/} {\bf 17}, 062101.

\bibitem[Zel'dovich \& Raizer(2002)]{Zeldovich}
{\sc Zel'dovich, I.A.B. \& Raizer, Y.P.} 2002 {\em Physics of Shock Waves and
  High-Temperature Hydrodynamic Phenomena\/}. Dover Publications.

\end{thebibliography}

\end{document}